# A strategy for a Kaluza-Klein graviton discovery at LHC

Alain Le Yaouanc[1], François Richard[2]

Université Paris-Saclay, CNRS/IN2P3, IJCLab, 91405 Orsay, France

October 12, 2026

**Abstract**

*Evidence for excesses observed at LHC tend to cluster mostly around 95 GeV and 650 GeV. We claim that a consequent interpretation of the 650 GeV and some additional indications calls for a Randall Sundrum sequence of Kaluza Klein graviton resonances. A common prejudice claims that such resonances should be very heavy, out of reach for LHC. We argue that this is no so. Although present LHC searches seem to exclude these particles below 2 TeV, we explain why this 650 GeV resonance could be one of them. We also explain why ongoing analyses ignore this scenario and apply selections which are only valid for scalar resonances. This could also apply to ongoing searches for Higgs pairs where the search for resonances decaying into SM Higgs pairs ignores the KK graviton scenario. We expect that with a correct treatment, one could very significantly increase the detection efficiency for ZZ and hh channels and achieve, with present data, a convincing proof of our expectations. Channels like h(95)h(125) and h(95)h(95) should be added to present selections. Such improvements should allow to reach a discovery. For the ZZ and WW channels, one can use semi-leptonic and purely hadronic final states, to confirm the tower of KK resonances indicated up to 3 TeV.*

**Follow up of discussions at the Higgs Hunting meeting 2026 in Orsay**

---

1 Alain Le Yaouanc <alain.le-yaouanc@ijclab.in2p3.fr>
2 François Richard <francois.richard@ijclab.in2p3.fr>

# I. A summary of BSM searches

In a recent compilation of indications for BSM physics, reference [1] concludes that two channels seem to be emerging :

- A large accumulation of indications for the ‘650 GeV’ which is interpreted by us in [2,3]
- Three indications for a resonances at 95 GeV interpreted in several references

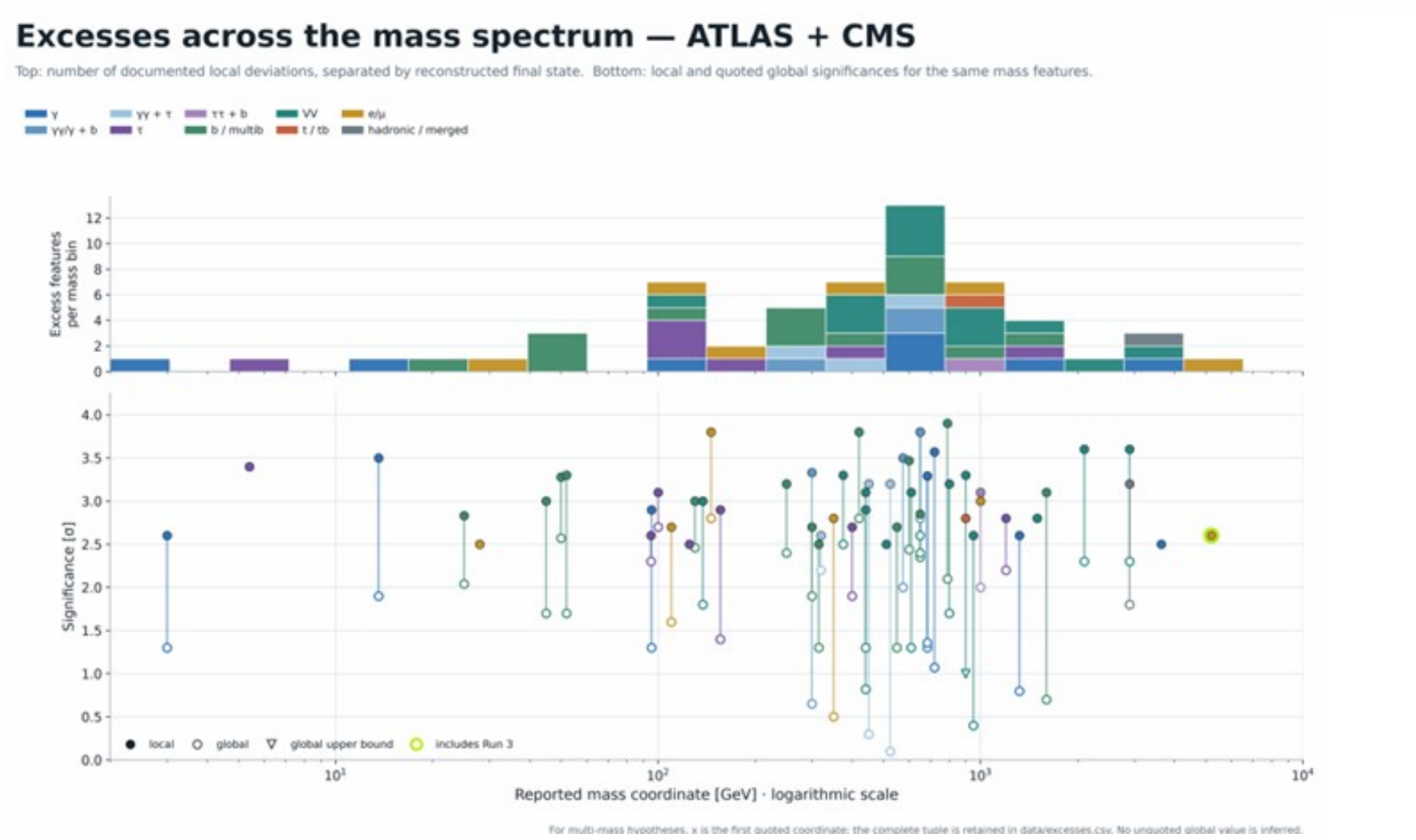

*Figure 1: From [1], list of excesses recorded by ATLAS and CMS*

Both efforts therefore converge towards the conclusion that LHC could soon reach a BSM discovery. Furthermore our interpretation provides a well defined theoretical framework which should help in avoiding mistakes and reach sooner this splendid goal. The table below summarises our findings.

| Channel | Local significance s.d. | Reference |
|---|---|---|
| T700→ZZ→$\ell^+\ell^-+\ell'^+\ell'^-$ | 3.5 ggF 2.7 VBF | ATLAS2009.14791 |
| T700→WW→$\ell^+\ell^-+\nu\nu$ | 3.8 VBF | CMS-PAS-HIG-20-016 |
| T700→h125h95 $b\bar{b}/\tau\tau+\gamma\gamma$ | 3.8 | CMS 2310.01643, 2506.23012 |
| T700 → $\gamma\gamma$ | 3.5 | ATLAS 2102.13405 |
| T700→e+e- | 3 | ATLAS1903.06248 CMS2103.02708 |
| H650+T700→Jet-Jet | 3.4 | ATLAS 2509.01219 |
| H650→A490Z→$t\,\bar{t}\,\ell^+\ell^-$ | 2.85 | ATLAS 2311.04033 |
| H650→T380h125→ $b\bar{b}b\bar{b}$ | 4.1 | CMS-PAS-HIG-20-012 |
| H650→$t\,\bar{t}$ | 3 | ATLAS 2404.18986 CMS 2507.05119 |

While no single channel satisfies the 5$\sigma$ discovery criterion, we claim that the combination of evidences for '650 GeV' indications reaches a global significance well above 5$\sigma$ and is supported by a consistent theoretical interpretation that will be developed below.

We believe that, on top of the T700 spin 2 candidate, it is likely that there is a scalar labelled H650. Mass resolution alone does not allow to separate them. This separation is intuitively based on the final states.

# II. A KK graviton candidate?

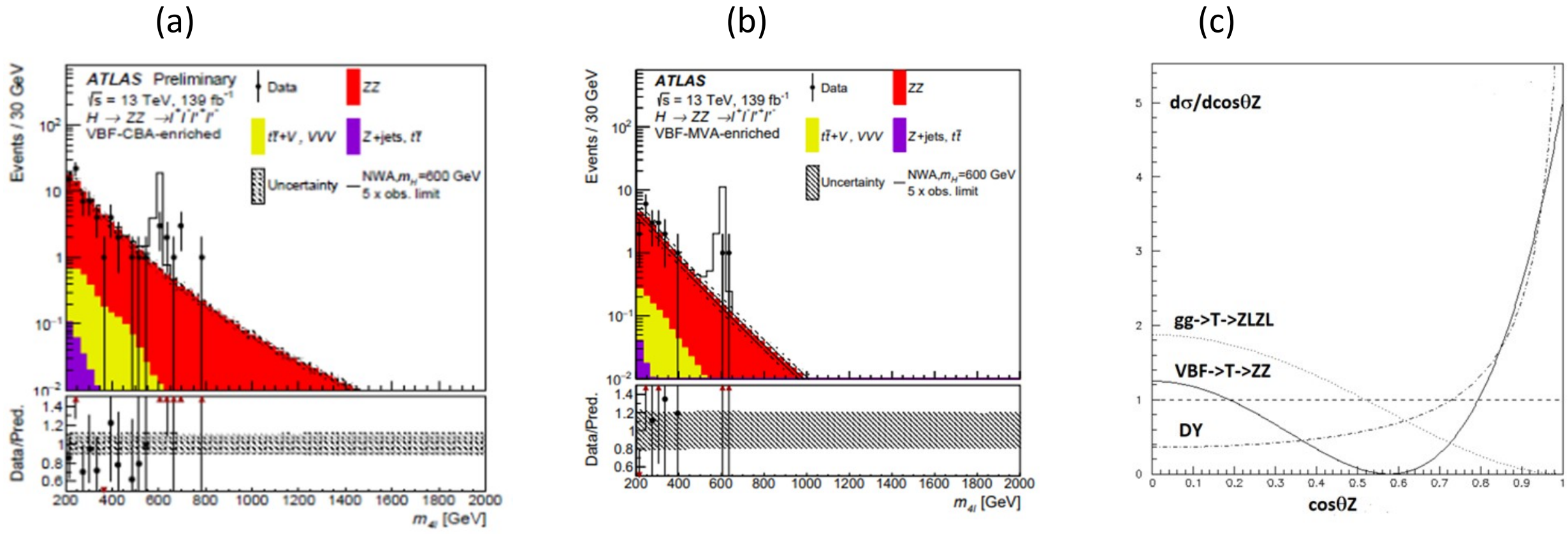

*Figure 2:(a) shows the excess observed in VBF in a genuine cut based analysis while (b) shows an almost disappearance of this excess applying a sophisticated selection against the Drell Yan background which should have no significant effect for a scalar resonance as shown by the simulation. (c) shows a forward peaking for VBF→T700→ZZ analogous to the DY distribution.*

Recall that our interpretation of T700 as a tensor (J=2) has been triggered by two analyses of the ZZ channel into four leptons by ATLAS shown in figure 2. The first analysis, a genuine cut approach, shows evidence for the '650 GeV' both in the ggF+VBF and the VBF channels (figure 2a). A sophisticated analysis optimised for a scalar resonance resulted in the disappearance of a large fraction of the candidates, only 2 remaining out of 9, as recalled in figure 2b. This can be easily understood by saying that cutting events with $|\cos\theta|>0.75$ efficiently removes the Drell Yan background while keeping 75% of a scalar signal, while it cuts ~2/3 of the signal for a J=2 resonance, as observed in figure 2b with strict scalar selection. CMS did not perform a cut based analysis.

Our interpretation of T700 in terms of a Kaluza Klein graviton tends to be swept away by some experts, arguing that such resonances are abundantly produced in the ggF mode, unaffected by the scalar selection, and should also be easily observed into photon pairs. This led to the hasty conclusion that such particles should be heavier than 2 TeV. This statement seemed backed up by the T and S, the Peskin and Takeuchi variables and by the SM Higgs BR, which leads to the idea that KK particles should be very heavy, beyond LHC present reach (see Appendix II). This conclusion was however attenuated by phenomenologists realising that, with extra symmetries, one could decrease the T,S limit to 2 TeV, noting also that this limit need not apply universally to all categories of KK particles.

By analogy with the SM Higgs case, we suggest that there is no direct coupling of these resonances to gluons and photons but only indirect couplings through top and W loops. This behaviour is interpreted in a dual composite description of our world, by saying that the SM Higgs is composed of colourless and charge-less preons. Why should this not happen with KK gravitons?

At this stage one is led to the conclusion that there could be such resonances with masses below a TeV, mainly produced through VBF, as explicitly stated in the CMS analysis for the WW channel. This statement may seem contradictory with the fact that the SM Higgs is abundantly produced through loops in the ggF process. This leads to the additional consequence that the coupling of these resonances to top quarks is unexpectedly weak. In the RS framework this can be accommodated by saying that the top quark bulk location weakly overlaps with the KK gravitons (see Appendix V of [3]).

In conclusion of these premisses, the main decay modes of these KK graviton resonances are WW/ZZ/hh, (recall that ZL is equivalent to h) with ratios 2/1/1, in accord with the RS model. So far only the first two modes were observed together with h95h125 for T700, while T1000 is observed into h125h125. This behaviour could come from the statistical factor which affects identical boson states and from a background decreasing with mass which allows to observe h125h125 only at higher masses.

# III. A tower of RS resonances?

Indications for T700 into 2 jets, 2 photons, e+e-, top pairs, X380h125 complete the list of indications. We claim that X380 is the first KK graviton of the RS sequence. It is indicated into ZZ with also an inadequate angular treatment and therefore needs confirmation. This resonance is altogether indicated in 5 modes [3].

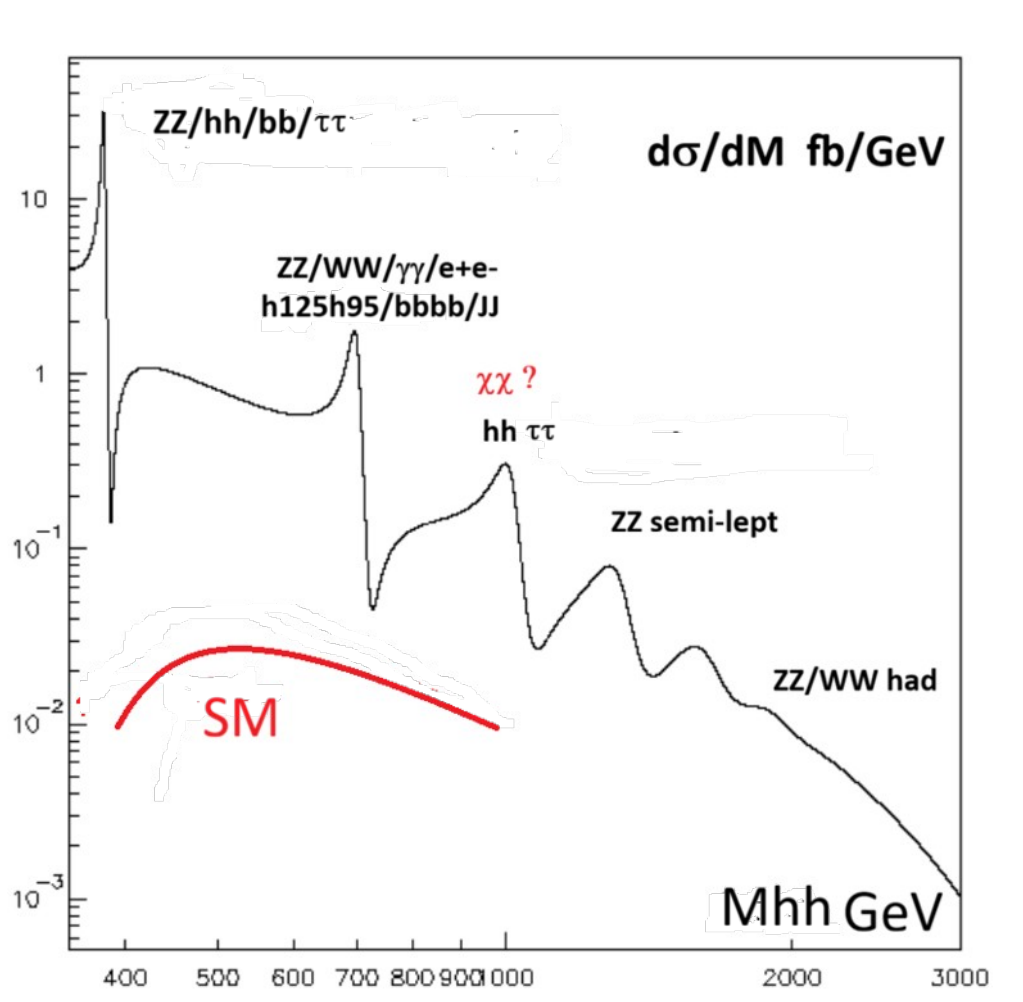

*Figure 3: Predicted VBF differential cross section for KK gravitons→hh. Various observed indications coincident with predicted KK graviton resonances are reported. In red is the SM expectation, an order of magnitude below.*

Given its importance for future machines, a few comments are of order about the e+e- indications provided by ATLAS and CMS. If confirmed, they would result into a fabulous opportunity to produce the T700 resonance and its satellites at e+e- colliders (see Appendix I). With a realistic integrated luminosity, one could achieve gigafactories for some of these resonances.

Several additional candidates to T700 and T380 are compatible with an RS interpretation (see Appendix I), at 1000 GeV in hh and ττ[2], 1300 GeV in ZZ decaying semi-leptonically[2], followed by a continuum of wide resonances seen in ZZ/WW hadronic modes[2].

We therefore think that the evidence for a spin 2 resonance, T700, plus indications for a tower of resonances satisfying the mass pattern predicted by RS (see Appendix I), reinforces our KK graviton interpretation and opens promising perspectives of discoveries which could be accomplished with RUN2+RUN3 data.

# IV. The Higgs pair sector

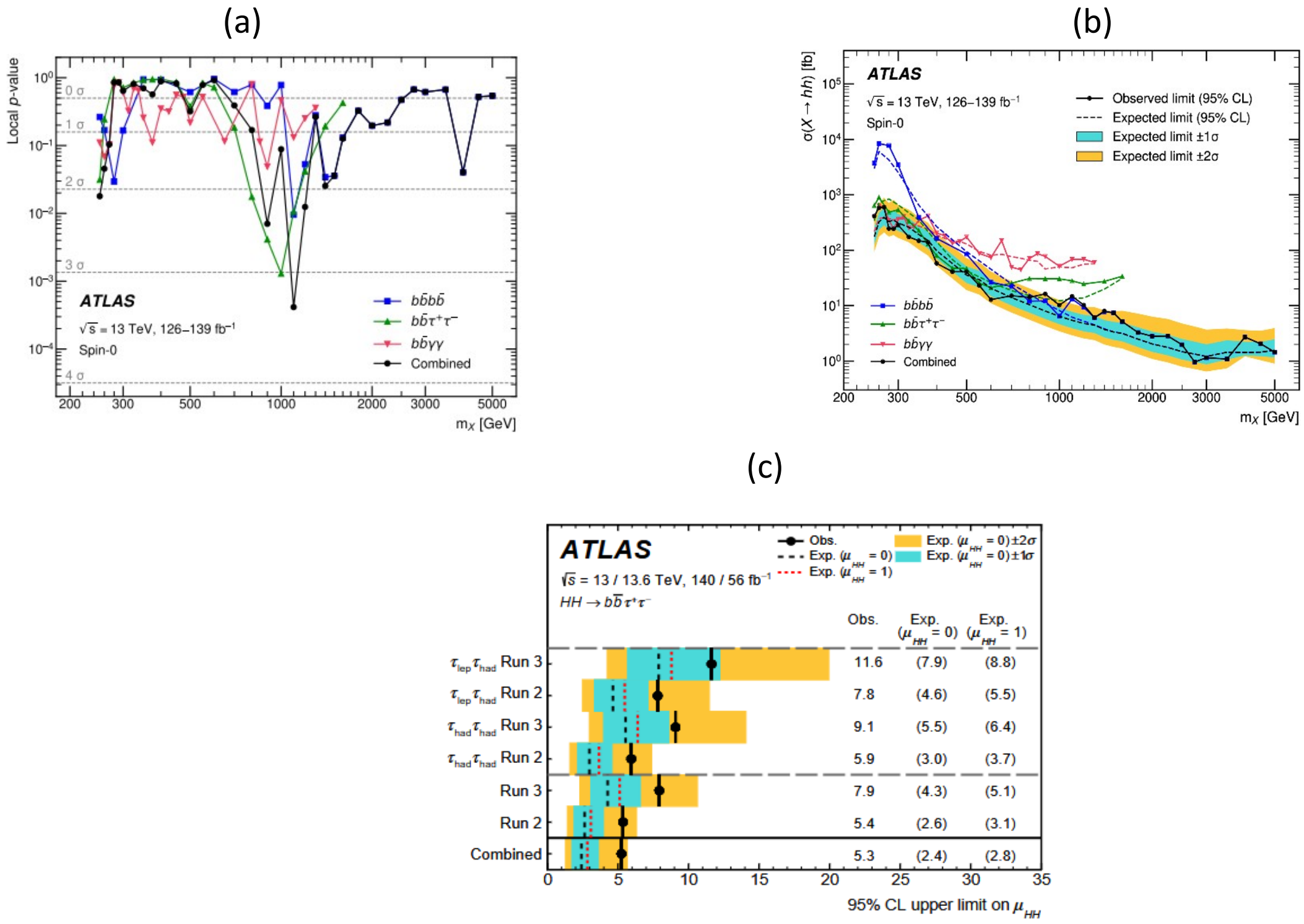

*Figure 4: Results for the hh analysis in ATLAS. (a) indicates an excess at 1.1 TeV, mainly coming from the $b\bar{b}\tau^+\tau^-$ mode as indicated by (b). (c) shows an excess in µHH which is also observed by RUN3.*

Evidence for T1000→hh shown in figure 4a has greatly benefited from the high priority given to the triple Higgs coupling measurement. So far these analyses obviously assume that there are no resonances decaying into hh, while figure 3 predict a major contribution from BSM resonances. The triple Higgs coupling generates isotropic decays which is not the case for the KK graviton resonances which could mean that the standard selections are inefficient for a J=2 resonance as it was already the case for the ZZ channel. While there is no explicit angular selection in the BDT algorithm of ATLAS[3], there could be such an effect, e.g. by using as an input the transverse momenta pT of the photons in the bbγγ mode since the h particles from VBF→T→ hh are distributed as the Z particles, therefore with reduced pT with respect to the scalar case.

Figure 4c shows the present situation at LHC which confirms an excess in the variable µHH but weaker than expected. CMS observes this excess only in the RUN3 analysis.

Some action should be taken to remedy this situation. In any case, one needs to fully understand the KK component which is likely to fake an apparent alteration of the SM Higgs potential.

As we said, there is some evidence that the ‘650’ decays into h(125)h(95)→bbγγ. This suggests that one could considerably increase the efficiency for these searches in scalar pairs by including h(95)h(95) and h(95)h(125). By how much, one cannot tell since we do not precisely know the BR of h(95) into two photons, ττ and bb modes.

3 The thank Marc Escalier for discussing this aspect of his analysis.

All in all we estimate that one could improve the efficiency for present searches by an order of magnitude and reach discovery.

# V. Preserving unitarity

Still marginal but promising are some excesses in the $W^+W^+$ and $ZW^+$ final states. The SM reaction $W^+_L W^-_L \rightarrow W^+_L W^-_L$ receives contributions from the three SM particles Z/photon/Higgs in the s and t channels. For large s values, these contributions cancel such that the resulting amplitude does not diverge, preserving unitarity at large s. This remains true adding extra scalar doublets, given that heavy neutral scalars decouple from WW/ZZ and do not contribute at large s.

Note in passing that this feature further rejects the WW/ZZ indications as coming from a heavy scalar belonging to an isodoublet. In this model, one predicts that if the lightest scalar h125 is maximally coupled to ZZ/WW – which is the case – the heaviest one, H650, being decoupled from ZZ/WW.

**T**he lightest KK graviton resonance at 380 GeV is likely to generate an early unitarity violation unless the s channel contribution is cancelled by **u channel** contributions coupled to ZW and $W^+W^+$ resonances. Evidence for these are provided by LHC experiments, as shown in figure 5. This would imply that massive KK gravitons belong to a I=2 representation, as considered in reference [4]

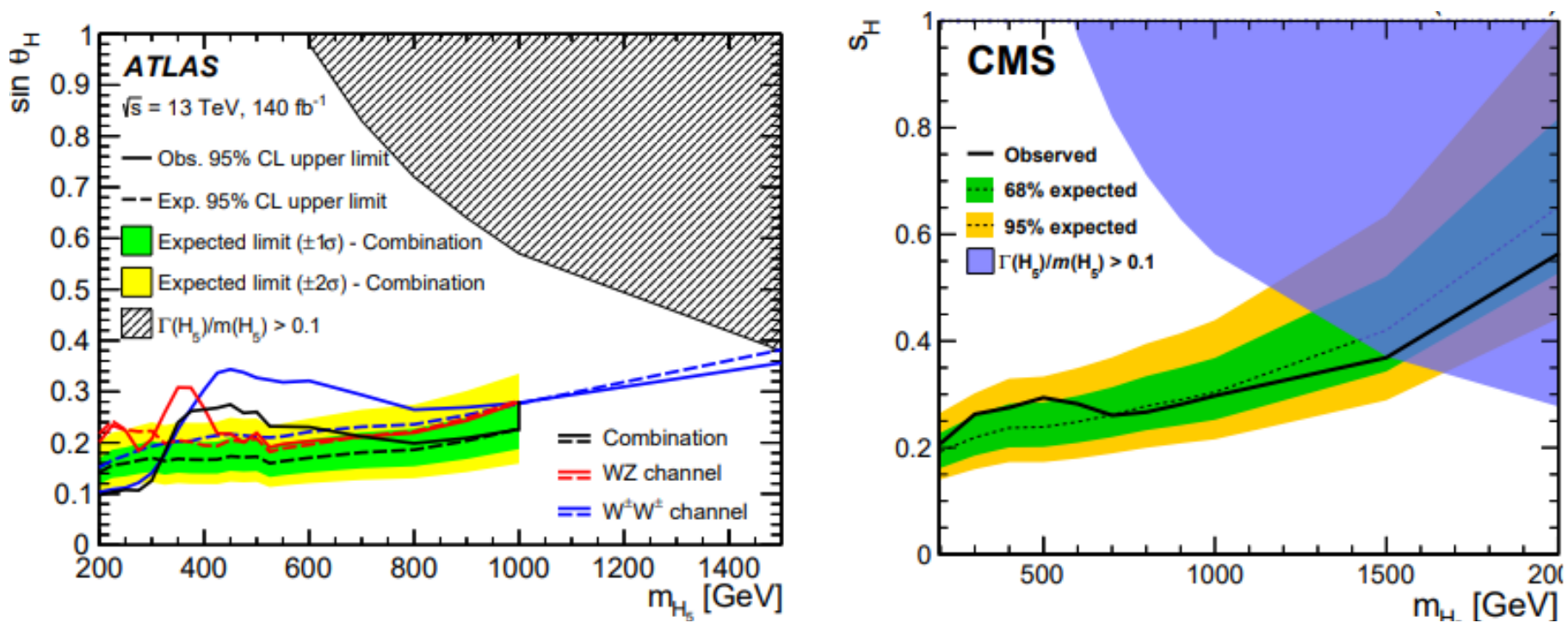

*(a)* *(b)*

*Figure 5: Evidence for an excess in the WZ and W+W+ modes in ATLAS (a) and CMS (b).*

# Conclusion

To conclude we think that, with the improvements proposed in this note, it would be possible to discover BSM physics at the present level of integrated luminosity reached by LHC, which would provide an invaluable asset for HEP physics in order to support an ambitious HEP project or, at least, trigger a new enthusiasm in LHC searches. We underline that there exist a consistent theoretical framework which allows to interpret a large set of experimental observations beyond the standard model.

# Postscriptum

Recently [5], we became aware of a possible connection between the Fermi-LAT indication for a dark matter candidate observed from the galactic halo but not from dwarf galaxies and the presence of these J=2 resonances. This effect can be understood given that the DM particles would annihilate into a p-wave, greatly suppressed in dwarf galaxies with velocities ten times smaller. This idea would not work with a scalar resonance.

This evidence is clearly well below those from LHC discussed in the present note, in spite of the recent appearance of a DM candidate with a similar mass in the LZ detector. The interested reader can consult [5]. If this DM is a SUSY neutralino, there would be excellent prospects to produce massively and measure very precisely charginos and neutralinos properties in gigafactories produced by KK graviton resonances.

**Acknowledgments**

*Many thanks to **Louis Fayard** for encouraging my participation to the **Higgs Hunting** meeting. Useful discussions at this meeting are gratefully acknowledged, in particular with **Livia Soffi, Marc Escalier and Laura Reina**. Stimulating discussions with **Jenny List** are kindly acknowledged.*

# APPENDIX

## I. SUMMARY OF NUMERICAL RESULTS

**ZZ/WW e+e- $\gamma\gamma$ hh $\tau\tau$ ZZ semi-lept VBF ZZ+WW had**

| m $G_{KK}$ GeV | 380 | **700±20** | 1000 | 1322 | 1640 | 1945 | 2260 | 2590 | 2905 |
|---|---|---|---|---|---|---|---|---|---|
| $\Gamma$ $G_{KK}$ GeV | 3 | **20±7** | 60 | 130 | 260 | 440 | 680 | 1042 | 1440 |
| VBF→$G_{KK}$ fb | 360 | **100±50** | 38 | 15 | 7 | 3.4 | 1.7 | 0.9 | 0.4 |
| e+e-→ $G_{KK}$ pb | 450 | **130** | 60 | 50 | | | | | |

*Table: Predictions for the KK graviton VBF cross sections and total widths. The line above indicates some observed final states for the various mass regions.*

In the RS model, one predicts a sequence of KK graviton resonances $G_{iKK}$ which have **quantified masses**. These masses are proportional to the **zeros of the Bessel function $J_1$(x):**

**xi=3.83,7.02,10.17,13.32,16.5,19.6,22.8,25.9,29.05,32.2,35.3…**

Present indications give **mi~100$x^G_i$**,in GeV**,** while the total width itself goes like **$\Gamma$i~0.06($x^G_i$)$^3$** in GeV.

So far we have candidates at 380 GeV (5), 700 GeV (9) , 1 TeV (2), 1.3 TeV(1) and a continuum of wide resonances observed in hadronic modes from WW/ZZ for higher masses up to 3.5 TeV [2,3].

For what concerns e+e- colliders one predicts the following pattern:

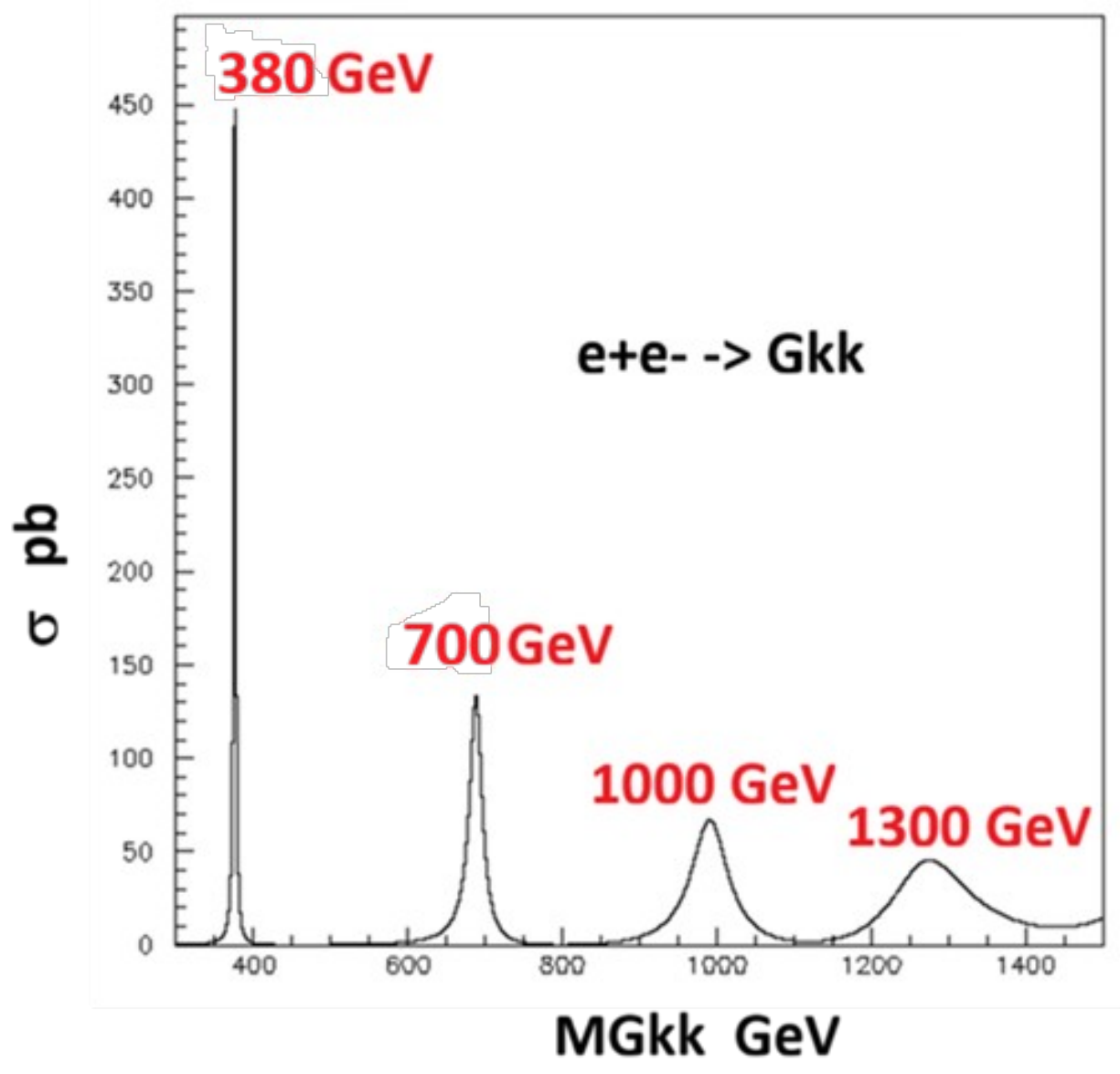

*Figure 5: Predicted e+e- cross sections for KK graviton resonances deduced from e+e- indications in ATLAS and CMS.*

Above cross sections predict gigafactories for some KK graviton resonances given that, for instance, ILC expects to deliver 8000 fb-1 at ECM=1 TeV.

This prediction heavily relies on the two indications provided by ATLAS and CMS shown on figure 6. A confirmation of these indication is therefore essential to demonstrate the great potential of an e+e- LC.

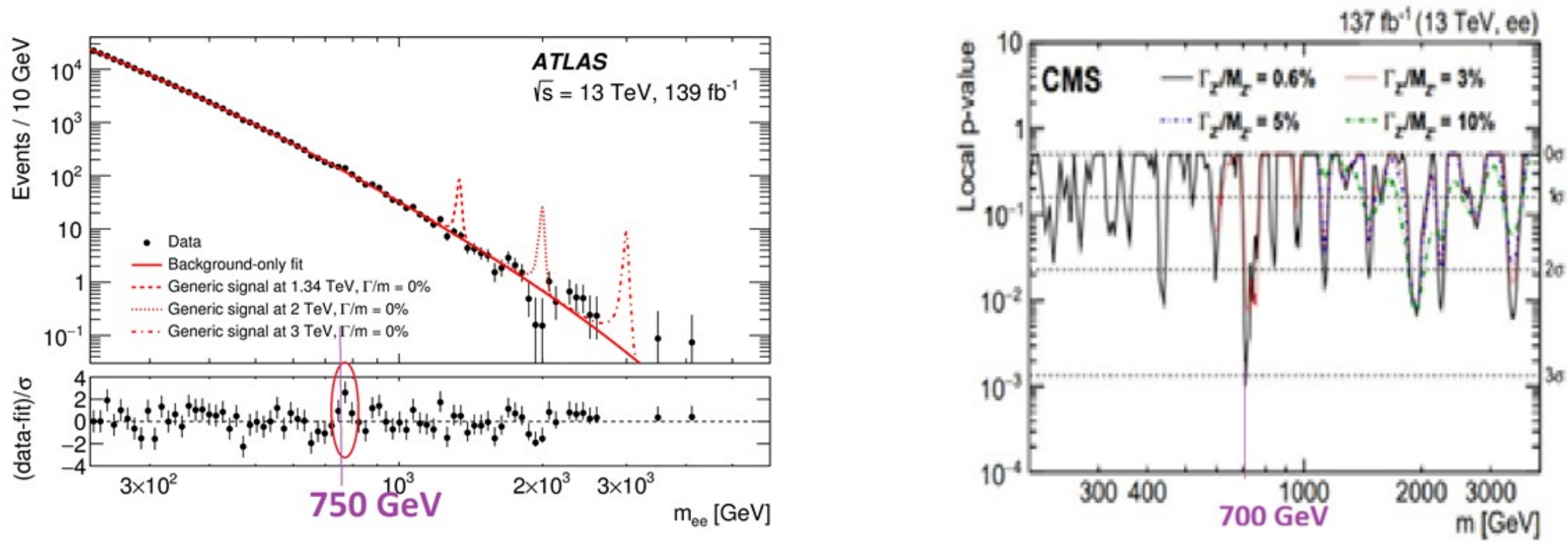

(a) (b)

*Figure 6: (a) e+e- mass distribution observed in ATLAS indicating a small excess around 750 GeV. (b) P-value distribution observed by CMS showing a narrow excess in the e+e- distribution at 700 GeV.[4]*

In the RS model, one expects that the electron coupling to KK gravitons will be very weak. This is probably true for eL which belongs to a weak doublet with a neutrino but, perhaps, not for eR. A LC with polarisation would immediately reveal this capital feature.

## II. PRECISION MEASUREMENT PREDICTIONS

In this section we ask ourselves whether precision measurements are compatible with the light KK graviton masses indicated by LHC data.

For the KK phenomenology, one defines a unique **mass scale parameter mKK**, not directly related to the mass of a given KK particle, meaning vector or fermion KK masses could be heavier than KK tensor masses. T and S constraints imply [6] that **mKK>30 TeV**. With **custodial symmetry** one has **mKK>1.8 TeV**. KK gravitons could be lighter. The custodial symmetry implies an extension of the SM with additional symmetries, with the **presence of a Z'** with a mass similar to the KK vectors which seems in contradiction with present Z' mass limits, meaning that universality of KK masses is badly broken.

Reference [7] suggests that breakdown of KK mass universality can be accomplished within the kinetic terms.

More recently this issue was re-examined in [8], with the conclusion that constraints from the flavor sector may push these mass limits to much higher scales, even beyond FCChh. It is unclear how this would affect the graviton sector.

If HL-LHC confirms our findings, this will certain help to clarify these various issues, perhaps opening the opportunity for a discovery of a Z' at LHC.

4 See 1903.06248 for ATLAS and 2103.02708 for CMS.